\documentclass[aps, prb, twocolumn, superscriptaddress , showpacs]{revtex4-1}

\usepackage{graphicx}
\usepackage{dcolumn}

\begin{document}

\title{Phase transformation in Si from the semiconducting diamond to
  the metallic $\beta$-Sn phase in QMC and DFT under hydrostatic and
  anisotropic stress.}

\author{R. G. Hennig}

\email{rhennig@cornell.edu}

\homepage{\url{http://www.ccmr.cornell.edu/~rhennig}}

\affiliation{Department of Materials Science and Engineering, Cornell
University, Ithaca, New York, 14853-1501, USA}

\affiliation{Department of Physics, The Ohio State University,
Columbus, Ohio 43210, USA}

\author{A. Wadehra}

\affiliation{Department of Physics, The Ohio State University,
Columbus, Ohio 43210, USA}

\author{K. P. Driver}
\affiliation{Department of Physics, The Ohio State University,
Columbus, Ohio 43210, USA}

\author{W. D. Parker}
\affiliation{Department of Physics, The Ohio State University,
Columbus, Ohio 43210, USA}

\author{C. J. Umrigar}
\affiliation{Laboratory of Atomic and Solid State Physics, Cornell University, Ithaca, New York 14853, USA}

\author{J. W. Wilkins}
\affiliation{Department of Physics, The Ohio State University, Columbus, Ohio 43210, USA}

\date{\today}

\begin{abstract} Silicon undergoes a phase transition from the semiconducting diamond phase to the metallic $\beta$-Sn phase under pressure. We use quantum Monte Carlo calculations to predict the transformation pressure and compare the results to density functional calculations employing the LDA, PBE, PW91, WC, AM05, PBEsol and HSE06 exchange-correlation functionals.  Diffusion Monte Carlo predicts a transition pressure of $14.0 \pm 1.0$ GPa slightly above the experimentally observed transition pressure range of 11.3 to 12.6~GPa.  The HSE06 hybrid functional predicts a transition pressure of 12.4~GPa in excellent agreement with experiments. Exchange-correlation functionals using the local-density approximation and generalized-gradient approximations result in transition pressures ranging from 3.5 to 10.0~GPa, well below the experimental values.  The transition pressure is sensitive to stress anisotropy.  Anisotropy in the stress along any of the cubic axes of the diamond phase of silicon lowers the equilibrium transition pressure and may explain the discrepancy between the various experimental values as well as the small overestimate of the quantum Monte Carlo transition pressure.\end{abstract}


\maketitle

\section{Introduction}

Phase transformations between insulating or semiconducting and metallic phases present a challenge for many current theoretical methods.  These transitions provide a testing ground for comparing the accuracy of quantum Monte Carlo (QMC) methods and novel exchange-correlation functionals used in density functional theory (DFT).  In this work, we investigate the diamond to $\beta$-Sn phase transition in silicon with QMC methods and with various types of  exchange-correlation functionals in DFT.

Under pressure, Si displays eleven phases with a steady increase in coordination number and a transition from semiconductor to metal.\cite{Mujica03} At ambient pressure, Si occurs in the diamond structure.  At a pressure of about 12~GPa, Si transforms to the high-pressure $\beta$-Sn structure.\cite{Jamieson63} This phase transition coincides with a semiconductor-to-metal transition and an increase in the coordination number from 4 to 6.  The transformation reduces the space group symmetry from cubic Fd$\bar 3$m (227) to tetragonal I4$_1$/amd (141).  At the transition pressure, the volume decreases by 21\%.\cite{Hu86, McMahon94} It is noteworthy that the melting of Si at ambient conditions also displays a transition both from fourfold to sixfold coordination and from the semiconducting solid phase to a metallic liquid.



The observed transition pressure in Si from the diamond to the $\beta$-Sn phase depends strongly on the experimental conditions and is affected by non-hydrostatic stresses.  Under nominally hydrostatic conditions in diamond anvil cell experiments using a pressure medium, the observed transition pressure ranges from 11.3 to 12.6~GPa.\cite{Piermarini75, Weinstein75, Welber75, Werner82, Olijnyk84, Hu86, McMahon94} Without a pressure medium or under uniaxial compression, Si transforms at significantly lower pressures of 8--9~GPa.\cite{Gupta80, Hu86} In shock compression, the transformation is observed in a range of 10--14~GPa.~\cite{Altshuler65, Pavlovskii68, Duvall77}


Previous calculations using DFT determined the phase transition pressure for various semilocal exchange-correlation functionals.  Comparing the results of zero-temperature calculations directly with room-temperature experiments requires taking into account the phonon contributions to the free energy.  Zero-point energy and finite temperature phonon entropy contributions to the free energy lower the transition pressure by 1.0 and 0.3~GPa, respectively.~\cite{Gaal-Nagy99} Taking this into account, the previous hydrostatic compression calculations gave transition pressures of 5.7--6.7~GPa using the local density approximation (LDA) and 10.1--10.9~GPa using the PW91 functional, a generalized gradient approximation (GGA) -- all below the experimental range.~\cite{Gaal-Nagy99, Moll95, Needs95, Cheng01} For non-hydrostatic compression, DFT calculations confirm the experimental observation that stress anisotropies can lower the transition pressure.  The transition pressure decreases linearly with increasing deviatory stresses along one of the cubic axes.~\cite{Cheng01, Gaal-Nagy06}

In this study, we perform QMC and DFT calculations to benchmark semilocal and hybrid exchange-correlation functionals for the diamond to $\beta$-Sn transformation in Si and determine how non-hydrostatic stress affects the transition pressure.  Section~\ref{sec:method} introduces the DFT and QMC methods. We analyze the accuracy of the various approximations in our QMC calculations to determine the accuracy of the transition pressure prediction.  Section~\ref{sec:results} compares the results of the QMC and DFT calculations for the transition pressure between the two Si phases with the experimental results.  The LDA~\cite{PW92} predicts a transition pressure that is too low.  The GGA functionals PBE~\cite{PBE} and PW91~\cite{PW91} improve the prediction but still underestimate the pressure.  The more recently developed GGA functionals WC~\cite{WC}, AM05~\cite{AM05} and PBEsol~\cite{PBEsol} perform worse than the PBE and PW91 functionals and predict transition pressures more similar to the LDA.  The hybrid functional HSE06~\cite{Heyd03, HSE06} gives a transition pressure of 12.4~GPa, in excellent agreement with the range of experimental values of 11.3-12.6~GPa.  Diffusion Monte Carlo (DMC) predicts a transition pressure of $14.0\pm 1.0$ GPa, which is slightly higher than the HSE06 results and the range of experimental values.  We discuss the origin of the shortcomings for the LDA and GGA functionals.  For the LDA, the lack of gradient terms of the functional results in an underestimation of the energy difference between the diamond phase and the more homogeneous $\beta$-Sn phase.  Based on a similar underestimation of the formation energies for interstitial defects in Si,~\cite{Batista06, Rinke09} we argue that the lack of nonlocal exchange in the LDA and semilocal GGA functionals is responsible for the remaining underestimation of the energy difference between the semiconducting diamond and $\beta$-Sn phases.~\cite{Batista06, Rinke09}

From the dependence of the energy on volume and $c/a$ ratio, we determine the effect of stress anisotropy on the transition pressure.  We show that stress anisotropies, which may be present under experimental loading conditions, lower the transition pressure and may explain the broad range of experimental values.  Section~\ref{sec:conclusion} summarizes the results.

\section{Methods}
\label{sec:method}

To accurately compare the predictions of various exchange-correlation functionals in DFT to QMC calculations, we have to either eliminate or control the physical and numerical approximations in both computational methods. For the DFT calculations we eliminate or control all approximations (other than the choice of the approximate exchange-correlation functional) to determine the transition pressure accurate to within 0.2~GPa.  In our QMC calculations we reduce the error introduced by the controlled approximations to result in an uncertainty of the transition pressure of 1~GPa. In the following, we describe our DFT and QMC approaches and estimate the accuracy of the approximations.

\subsection{Density functional methods}

\begin{figure}[tbh]
  \centering
  \includegraphics[width=8cm]{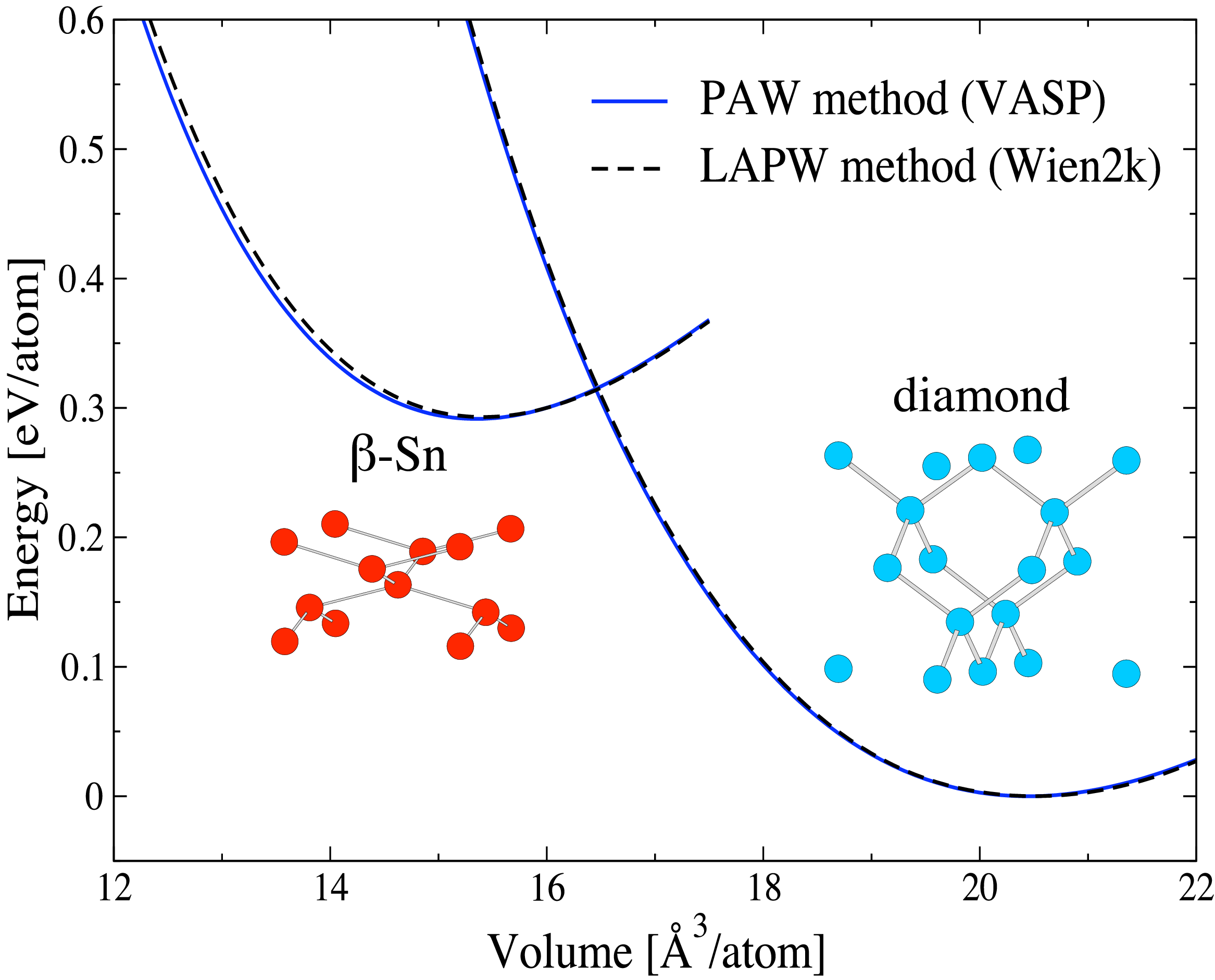}
  \caption{(color online) Accuracy of the frozen-core PAW method.  The energy of the diamond and $\beta$-Sn phases of silicon are shown for the all-electron LAPW method (Wien2k) and the frozen core PAW method (VASP) using the PBE functional .  The energies are shifted such that the minimum energy of the diamond phase is at zero.  The differences in energy predicted by the two methods at any given volume are within 5~meV/atom for the diamond phase and within 10~meV/atom for the $\beta$-Sn phase.  The transition pressures agree within 0.2~GPa.}
  \label{fig:comparisonPBE}
\end{figure}

The three main approximations for the density functional calculations, besides the choice of exchange-correlation functional, are the description of the core electrons, the accuracy of the basis set and the Brillouin zone integration.  For DFT calculations using semilocal exchange-correlation functionals, we explicitly include the core-electrons in the calculation and avoid the pseudopotential approximation by performing all-electron calculations using the LAPW method implemented in the Wien2k code.~\cite{Wien2k}  The parameters of the LAPW basis and the Brillouin zone integration are chosen to achieve a total energy accuracy of 1~meV/atom.  This requires a muffin-tin radius $R_\mathrm{MT} = 1.9\, a_0$ and a value of $R_\mathrm{MT}\,k_\mathrm{max} = 8.0$, where $k_\mathrm{max}$ is the planewave cutoff of the basis.  The Brillouin zone integration is performed on a $9\times 9 \times 9$ $k$-point mesh for the semiconducting diamond phase and a $15\times 15 \times 15$ $k$-point mesh for the metallic $\beta$-Sn phase.  The calculations for the hybrid functional HSE06 are performed using the VASP program (Vienna ab-initio simulation program) employing the PAW method within the frozen core approximation.~\cite{Kresse96, Kresse99}  A cutoff energy of 500~eV and a $15 \times 15 \times 15$ $k$-point mesh ensure convergence of the total energy to 1~meV/atom.

We test the accuracy of the frozen core PAW approximation by comparing calculations for the PBE functional with the all-electron LAPW method.  Figure~\ref{fig:comparisonPBE} illustrates the close agreement between the frozen core PAW and the LAPW method for the diamond and $\beta$-Sn phases of silicon.  At any given volume, the two methods agree to better than 10~meV/atom and the transition pressures agree to within 0.2~GPa.

\subsection{Quantum Monte Carlo methods}

There are two forms of QMC methods that are commonly used for electronic structure calculations, the simpler variational Monte Carlo (VMC) and the more sophisticated diffusion Monte Carlo (DMC) method.~\cite{Foulkes01, NigUmr-BOOK-99} VMC calculates quantum mechanical expectation values using Monte Carlo techniques to evaluate the many-dimensional integrals.  Accuracy of the VMC results depends crucially on the quality of the trial wave function, but DMC can remove most of the error in the trial wave function. DMC is a stochastic projector method that projects out the ground state from the trial wave function.  To avoid the fermion sign problem, one typically imposes the boundary condition that the nodes of the many-body wave function are the same as those of a trial wave function.  The resulting error, known as the fixed-node error, can be reduced by improving the trial wave function~\cite{Rios06, Umrigar07}

The QMC calculations are performed using the CHAMP code.~\cite{champ} A norm-conserving Hartree-Fock pseudopotential eliminates the 1$s$, 2$s$ and 2$p$ electrons of Si from the calculation.~\cite{Trail05} The QMC trial wave function consists of a product of a single Slater determinant of DFT orbitals and a Jastrow correlation factor. The orbitals of the Slater determinant come from a DFT calculation with the CPW2000 code of J.-L. Martins using the PBE functional.  For the diamond phase the L-point was chosen to reduce the finite-size error.  For the $\beta$-Sn phase the $(1/2, 0, 0)$ $k$-point was selected to reduce the finite-size error and to avoid any fractional occupancies of the orbitals.  The Jastrow factor describes the electron-electron and electron-nuclei correlations.  Minimizing the energy in VMC optimizes the parameters of the Jastrow.~\cite{Umrigar05, Umrigar07} Finally, DMC calculations using the optimized trial wave function determine the energy of the phases.

Approximations in the QMC calculations can be classified into controlled approximations with systematically-reducible error and uncontrolled approximations whose errors are unknown and are not systematically reducible. 

\begin{table}
  \caption{\label{tab:ControlledApproximations} Convergence of controlled approximations in QMC.  The parameters of the calculations are chosen to reduce the errors introduced by the controlled approximations to below 20~meV/atom.}
  \begin{ruledtabular}
    \begin{tabular}{l c l}
      Controlled & Convergence & Parameter \\
      approximation & [meV/atom] & value \\
      \hline
      Statistical error  & $<$3  & 30,000
      steps/$N_\mathrm{atom}$ \\
      Orbital interpolation grid & $<$ 5 &
      4.5--6.5~points/\AA \\
      DMC population control & $\ll$1 & 1,000 walkers \\
      DMC time step      & $<$ 5 & $\tau = 0.025$~Ha$^{-1}$ \\
      Finite size        & $<$ 20 & up to 128 atoms \\
    \end{tabular}
  \end{ruledtabular}
\end{table}

\begin{figure}[tbh]
  \centering
  \includegraphics[width=8cm]{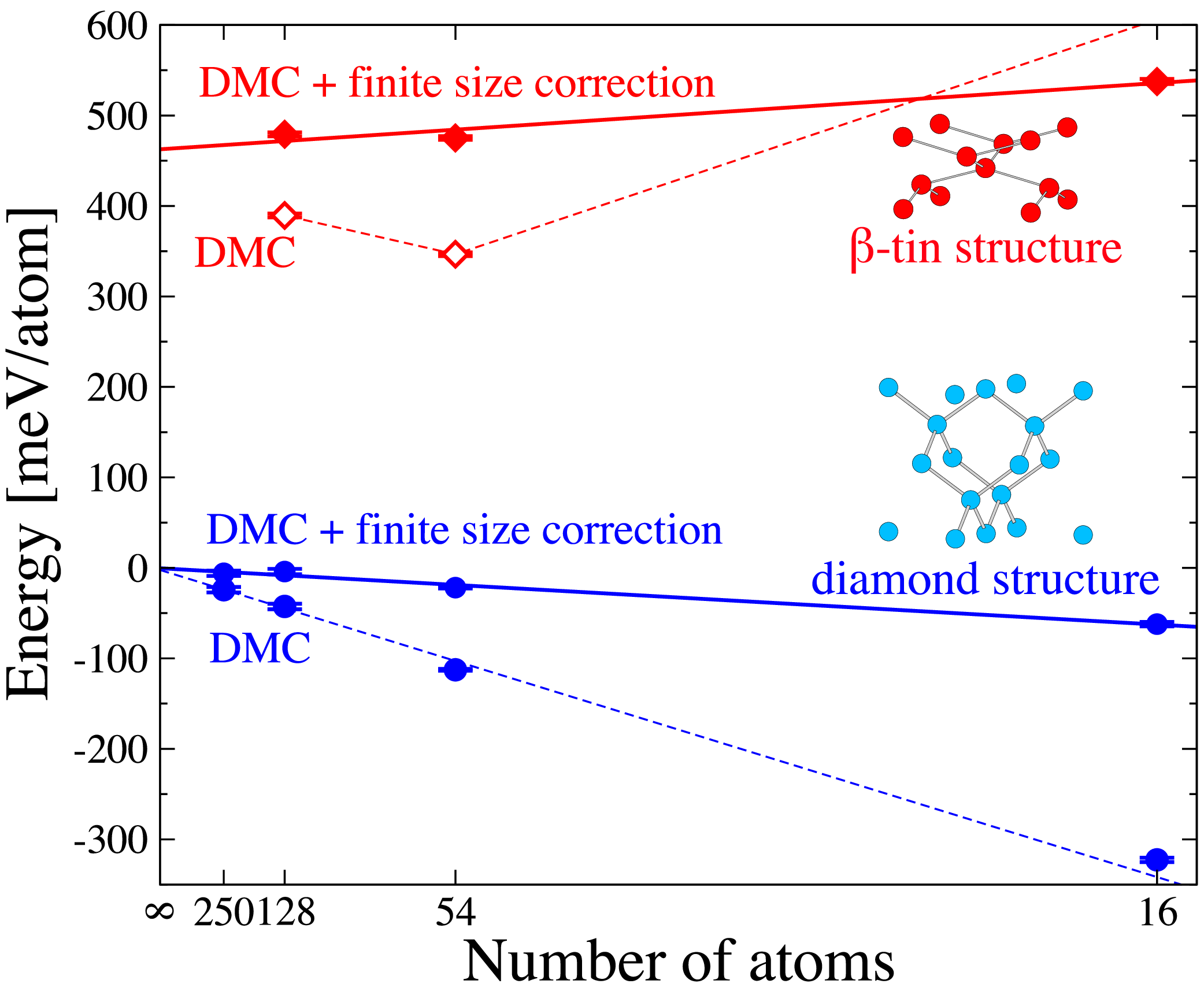}
  \caption{(color online) Finite-size extrapolation of the DMC energies for the diamond and $\beta$-Sn phases of Si.  The zero of the energy is taken to be the extrapolated value for the diamond structure including finite-size corrections.  The finite-size corrections using the finite-size exchange-correlation functional method~\cite{Kwee08} greatly reduce the finite-size error and enable an accurate finite-size extrapolation using the corrected DMC energies of the 16, 54 and 128 atom cells.}
  \label{fig:finsize}
\end{figure}

\emph{Convergence of the controlled approximations.}  Controlled approximations include the statistical error of the Monte Carlo method, the interpolation grid for the numerical orbitals, the system size, the number of configurations (walkers), and the time step in the diffusion Monte Carlo calculation.  Table~\ref{tab:ControlledApproximations} summarizes the accuracy of the controlled approximations in the QMC calculations.   All the errors introduced by the controlled approximations are reduced such that the resulting transition pressure is accurate to within 1~GPa, corresponding to an energy accuracy of about 20~meV/atom.

QMC calculations for extended systems require a finite-size extrapolation. The periodic boundary conditions employed in the calculation lead to artificial correlations of the electrons.  Several methods have been developed to reduce the finite-size error such as the model periodic Coulomb potential,~\cite{Kent99} the completion of the structure factor~\cite{Chiesa06} and the finite-size exchange-correlation functional approach.~\cite{Kwee08} Here we use the last approach, which employs a local density functional fit to DMC calculations for finite system sizes.~\cite{Ceperley80} The finite size corrections are evaluated using the implementation of the finite-size exchange-correlation functional in the PWSCF code.~\cite{PWSCF}

Figure~\ref{fig:finsize} shows the DMC energies for the diamond and $\beta$-Sn structure of Si at their respective equilibrium volumes as a function of the number of atoms.  Extrapolation from the raw DMC energies is difficult at best.  Using the finite-size exchange-correlation functional correction greatly improves the extrapolation of the energy to infinite system size, particularly for the $\beta$-Sn phase. ~\cite{Kwee08}  Using calculations with up to 128 atoms in the unit cell results in an extrapolation error below 20~meV/atom leading to an uncertainty in the predicted transformation pressure of about 1~GPa.

\begin{figure}[tbh]
  \centering
  \includegraphics[width=8.5cm]{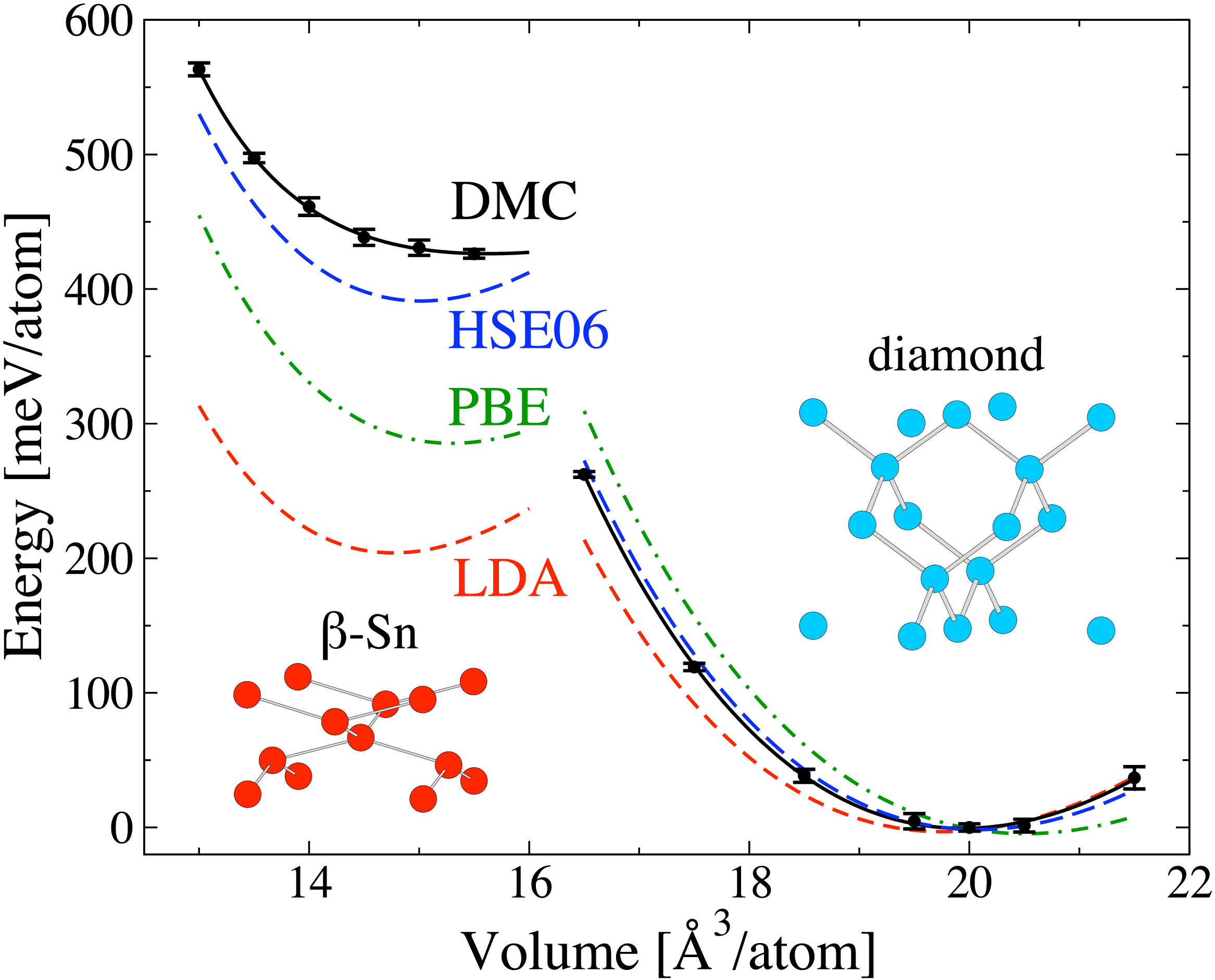}
  \caption{(color online) Comparison of the predicted equations of state of silicon for diffusion Monte Carlo energies and the LDA, PBE and HSE06 exchange-correlation functionals.  The energy is given relative to the energy of the diamond phase at a volume of 20~\AA$^3$. The DMC energy curve includes the core-polarization correction of 30~meV/atom described in Ref.~\onlinecite{Alfe04a}.}
  \label{fig:comparisoneos}
\end{figure}

\emph{Uncontrolled approximations.}  The uncontrolled approximations in QMC include the pseudopotential approximation for the core electrons, the locality approximation for the evaluation of the nonlocal pseudopotential terms and the fixed-node approximation in DMC.

Alf{\`e} {\it et  al.}~\cite{Alfe04a} showed that core-polarization, which is not included in our QMC calculations, is important for the energy difference between the $\beta$-Sn and the diamond phase and lowers the energy difference by 30~meV/atom.  To account for this effect, we apply a constant energy shift of 30~meV/atom to the DMC energy of the $\beta$-Sn phase.

In DMC, the trial wave function is used to evaluate the non-local part of the pseudopotential.  This pseudopotential locality approximation leads to a non-variational error in the DMC energies.  Empirically, this error introduced by the pseudopotential locality approximation is usually quite small for well-optimized trial wave functions.~\cite{Pozzo08}

The fixed-node error is difficult to estimate and could affect the results of our calculation.  Possible methods to improve the nodes of the trial wave function include orbital optimization~\cite{TouUmr-JCP-07,TouUmr-JCP-08} and backflow transformation.~\cite{Rios06} Both approaches are computationally very demanding and beyond the scope of this work.  Recent DMC calculations for self-interstitials in Si~\cite{Parker10} provide an estimate of the size of fixed-node error.  These calculations show that the backflow transformation significantly improves the accuracy of the trial wave function; the variance is reduced threefold by the backflow transformation.  The total energy for the diamond Si phase is only reduced by 12~meV/atom.  Furthermore, a partial error cancellation is observed for differences in total energy between the perfect bulk and the defects.  We expect a similar partial error cancellation of the fixed-node error between the diamond and the $\beta$-Sn phase.

\begin{figure}[tbh]
  \centering
  \includegraphics[width=8cm]{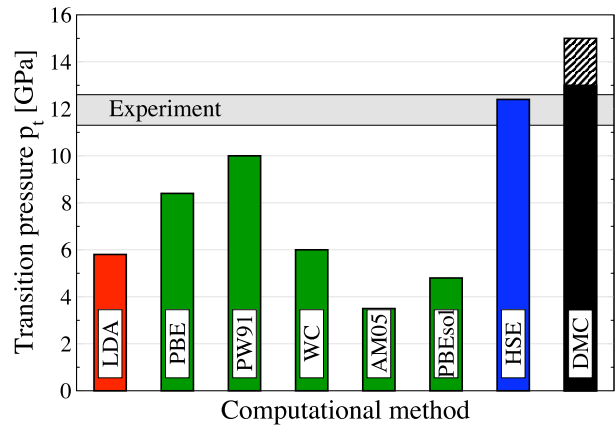}
  \caption{(color online) Comparison of predictions of the transition pressure for various exchange correlation functionals with DMC and experiment.  The LDA approximation (red) and all tested GGA approximations (green), underestimate the transition pressure.  The HSE06 hybrid functional (blue) agrees with the experiment. DMC (black) slightly overestimates the transition pressure.  The shaded region indicates the the estimate of the uncertainty of the DMC transition pressure from the controlled approximations.}
  \label{fig:comparisonpt}
\end{figure}

\section{Results}
\label{sec:results}

\subsection{Comparison between quantum Monte Carlo and various exchange-correlation
  functionals}

Figure~\ref{fig:comparisoneos} compares the predictions of the different exchange-correlation functionals for the energy as a function of volume of the diamond and $\beta$-Sn phases of silicon with the results of the diffusion Monte Carlo calculations.   For clarity, we show only the results for the LDA, PBE and HSE06 functional.

Table~\ref{tab:comparison} presents the equilibrium volume $V_0$, bulk modulus $B$ and pressure dependence of the bulk modulus $B'$ for Si in the diamond and $\beta$-Sn phases for the various theoretical methods.  For the diamond phase, we fit a Birch-Murnaghan equation of state~\cite{Birch47} to the energy as function of volume over a range from 16.0 to 22.0~\AA$^3$/atom.  Comparing the results with experiment shows an excellent agreement of the structural properties with experimental values for the more recent GGA functionals, WC, AM05 and PBEsol, and the hybrid functional HSE06.  The LDA, as usual, underestimates the volume, and the GGA functionals PBE and PW91 overestimate it.  A similar agreement is also observed for the volume $V_t$ at the experimental transition pressure of 11.7~GPa of Ref.~\onlinecite{McMahon94}.  All semilocal functionals somewhat underestimate the bulk modulus with the largest discrepancy observed for the PBE and PW91 functional.  All functionals reproduce the pressure dependence of the bulk modulus very well.  Within statistical accuracy, the DMC results agree with the experimental values for the diamond phase, especially for the equilibrium volume $V_0$.  The somewhat large error bars on $B$ and $B'$ make a comparison difficult.

\begin{table*}[t]
\begin{ruledtabular}
  \caption{\label{tab:comparison}Equation of state of the diamond and $\beta$-Sn phases of silicon in DMC and DFT with various functionals.  Shown are the equilibrium volume, $V_0$,  the bulk modulus, $B$, and the pressure dependence of the bulk modulus, $B'$ for both phases and the $c_0/a_0$ ratio for the $\beta$-Sn phase.  For comparison with experiment, we list the volume, $V_t$ and the $c_t/a_t$ ratio at the experimental transition pressure of 11.7~GPa of Ref.~\onlinecite{McMahon94}.   The experimental equilibrium volume of the diamond phase and the elastic properties are from Ref.~\onlinecite{Hull99}.  In addition, $\Delta E_0$, denotes the energy difference between the minima of the two phases.  The transition pressure, $p_t$ includes zero-point and finite-temperature corrections which lower the transition pressure by 1.0 and 0.3~GPa, respectively.~\cite{Gaal-Nagy99} The QMC transition pressure also includes a core-polarization correction which lowers the energy difference between the diamond and $\beta$-Sn phase by 30~meV/atom.~\cite{Alfe04a}}
  \label{tab:eos}  \begin{tabular}[t]{l c c c c c c c c c}
                                         &  LDA  &  PBE  &  PW91 &   WC  &  AM05 & PBEsol&  HSE06 &   DMC   & Exp. \\
    \hline
    \multicolumn{10}{c}{--- diamond phase --- } \\
    $V_0$ [\AA$^3$/atom]                 & 19.72 & 20.48 & 20.45 & 20.05 & 20.07 & 20.06 & 20.07 & 19.98(5) & 20.0 \\
    $B$ [GPa]                            & 96.4  & 89.0  & 89.1  & 94. 2 & 93.1  & 93.9  & 99.1  & 98(7)  & 99.2 \\
    $B'$                                 &  4.13 &  4.12 &  4.14 &  4.10 & 4 .08 &  4.09 &  4.00  &  4.6(6) & 4.11 \\
    $V_t$ [\AA$^3$/atom]\footnotemark[1] & 17.86 & 18.42 & 18.40 & 18.12 & 18.13 & 18.13 & 18.21 & 18.14(5) & 18.15 \\[0.5em]
    \multicolumn{10}{c}{--- $\beta$-Sn phase --- } \\
    $V_0$ [\AA$^3$/atom]                 & 14.82 & 15.36 & 15.47 & 15.11 & 14.82 & 15.02 & 15.10 & 15.2(1) \\
    $c_0/a_0$                            & 0.548 & 0.550 & 0.551 & 0.549 & 0.546 & 0.548 & 0.565 & 0.550\footnotemark[2] & \\
    $B$ [GPa]                            & 116.0 & 106.4 & 103.6 & 112.4 & 120.5 & 115.0 &  117.0 & 107(12) \\
    $B'$                                 &  4.59 &  4.57 &  4.52 &  4.41 &  4.54 &  4.52 &  4.35 & 4.6\footnotemark[3]  \\
    $V_t$ [\AA$^3$/atom]\footnotemark[1] & 13.63 & 14.04 & 14.11 & 13.86 & 13.67 & 13.80 & 13.89 & 13.9(1) & 13.96 \\
    $c_t/a_t$\footnotemark[1]            & 0.544 & 0.544 & 0.546 & 0.544 & 0.543 & 0.543 & 0.557 & 0.550\footnotemark[2] & 0.550\\[0.5em]
    \multicolumn{10}{c}{--- Phase transition --- } \\
    $\Delta E_0$ [meV/atom]              &  206  &  287  &  324  &  214  &  152  &  184  &  390  & 424(20) \\
    $\Delta V/V_0$ [\%]                  & 22.6  & 22.2  & 21.4  & 22.7  & 24.7  & 23.4  & 21.6  &    20.8(1)   &  21.0(1) \\
    $p_t$ [GPa]                            &  5.8  &  8.4  &  10.0 &  6.0  &  3.5  &  4.8  & 12.4  & $14.0\pm 1.0$ & 11.3-12.6 \\
\end{tabular}
  \footnotetext[1]{At the experimental transition pressure of Ref.~\onlinecite{McMahon94}.}
  \footnotetext[2]{The $c/a$ ratio is not optimized in DMC and fixed at the experimental $c/a = 0.550$.}
  \footnotetext[3]{A value of $B'=4.6$ is assumed in the DMC Birch-Murnaghan equation of state.}
\end{ruledtabular}
\end{table*}

For the $\beta$-Sn phase, we calculate the energy as a function of volume $V$ and $c/a$ ratio.  For each volume, we fit a cubic polynomial to determine the minimum energy and corresponding $c/a$ ratio.  Similar to the analysis for the diamond phase, a Birch-Murnaghan equation of state~\cite{Birch47} is then fit to the resulting energies as a function of volume ranging from 13.2 to 17.1~\AA$^3$/atom.  Since the $\beta$-Sn phase transforms under decompression into the R8 phase, no experimental values are available for the equilibrium crystal structure.  Instead we compare the volume and $c/a$ ratio at the experimental transition pressure of 11.7~GPa of Ref.~\onlinecite{McMahon94}.  The LDA and AM05 functional underestimate the volume at that pressure while all other functionals agree quite well with the experiment.

The transition pressure $p_t$ is determined from the equation of state using the common-tangent rule.  To compare the calculated bare transition pressure with experimental values, we need to include the effect of zero-point vibrations and finite temperature, which differ in the two phases.~\cite{Gaal-Nagy99}   The results for $p_t$ shown in Table~\ref{tab:comparison} include zero-point and finite-temperature corrections, which lower the transition pressure by 1.0 and 0.3~GPa, respectively.~\cite{Gaal-Nagy99}   Figure~\ref{fig:comparisonpt} compares the predictions for the transition pressure among the various exchange-correlation functionals and DMC with experiments.  Somewhat surprisingly, all semilocal functionals underestimate the transition pressure, and the GGA functionals provide a large range of pressure predictions from 3.5 to 10.0~GPa.  The PW91 functional predicts a pressure of 10.0~GPa close to the experimental range from 11.3 to 12.6~GPa.  The more recent GGA functionals, WC, AM05 and PBEsol, significantly underestimate the transition pressure, similarly to the LDA. The hybrid functional HSE06, however, predicts a transition pressure of 12.4 GPa, in excellent agreement with experiments.

The DMC calculations predict a transition pressure of $14.0 \pm 1.0$~GPa.  This value includes the same zero-point energy and finite-temperature phonon entropy contributions to the free energy~\cite{Gaal-Nagy99} as the DFT values, and the core-polarization correction by Alf\`e {\it et al.}~\cite{Alfe04a}.   The error bar of 1~GPa is an estimate of the combined uncertainties of the controlled approximations.

Our DMC transition pressure of $14.0 \pm 1.0$~GPa is slightly above the experimental range from 11.3 to 12.6 GPa.  The result is lower by 2.5~GPa than the DMC result of Alf\`e {\it et al.}~\cite{Alfe04a} of 16.5~GPa and agrees within error bar with the AFQMC result by Purwanto {\it et al.}~\cite{Purwanto09} of $12.6 \pm 0.3$~GPa.  The latter work estimated the transition pressure from the energy difference at the experimental transition pressure, limiting any estimate of the accuracy of the AFQMC method to the pressure value.

The work by Alf\`e {\it et al.}~\cite{Alfe04a} and our study determined the equation of state of both phases.  Both DMC calculations accurately predict the structural and elastic properties of the diamond phase demonstrating their accuracy.  The two DMC studies differ in their transition pressure prediction.  The higher transition pressure predicted by Alf\`e {\it et al.}  originates from a larger energy difference of $475 \pm 10$~meV/atom~\cite{Alfe04a} versus $424 \pm 20$~meV/atom in our work (both numbers include the core-polarization correction).  The difference of 2.5~GPa or 50~meV/atom between the two DMC results could be caused by differences in the pseudopotential locality error, the finite-size extrapolation or the fixed-node error.  The main differences in our approach are the use of energy minimization for optimizing the Jastrow factor parameters, a different form of the Jastrow and the finite-size extrapolation.  An improved optimization of the Jastrow parameters by the energy minimization method~\cite{Umrigar05, Umrigar07} factor would reduce the error of many of the controlled approximations.  It would also reduce the uncontrolled pseudopotential locality error in DMC.  In our work we include a finite-size correction~\cite{Kwee08} and perform a finite-size extrapolation at each volume for the two phases using the DMC energies for 16, 54 and 128 atom cells while Alf{\`e} {\it et al.} use the DMC energy of the 128 atom cells without extrapolation and finite-size corrections.  The combination of pseudopotential locality error and different finite-size extrapolation could explain the change in the energy difference of 50~meV/atom.

\subsection{Discussion of the phase stability}

The differences in the transition pressure predictions of the various methods are mostly determined by the relative phase stability or energy difference between the two phases.  The good agreement of the HSE06 hybrid functional with the experimental transition pressure indicates that the energy difference should be around 390~meV/atom.  The semilocal functionals all give energy differences that are too small, while the DMC energy difference appears to be a bit too large.  In the following, we discuss two arguments for the observed ordering of the energy difference between the semiconducting fourfold coordinated diamond phase and the metallic sixfold coordinated $\beta$-Sn phase for the different functionals.

We first show that the lack of gradient terms in the LDA functional~\cite{Ceperley80, Perdew81} results in an underestimate of the energy difference between the two phases.  Then, we argue that in order to accurately predict the energy difference between the two phases, the method also needs to predict the band gap accurately.  Both LDA and GGA functionals fail for the band gap.  The inclusion of exact exchange in the HSE06 functional recovers the band gap by improving the derivative discontinuity of the Kohn-Sham potential for integer electron numbers.~\cite{Perdew83, Becke93, Cohen08}

\begin{figure}[tbh]
  \centering
  \includegraphics[width=7.5cm]{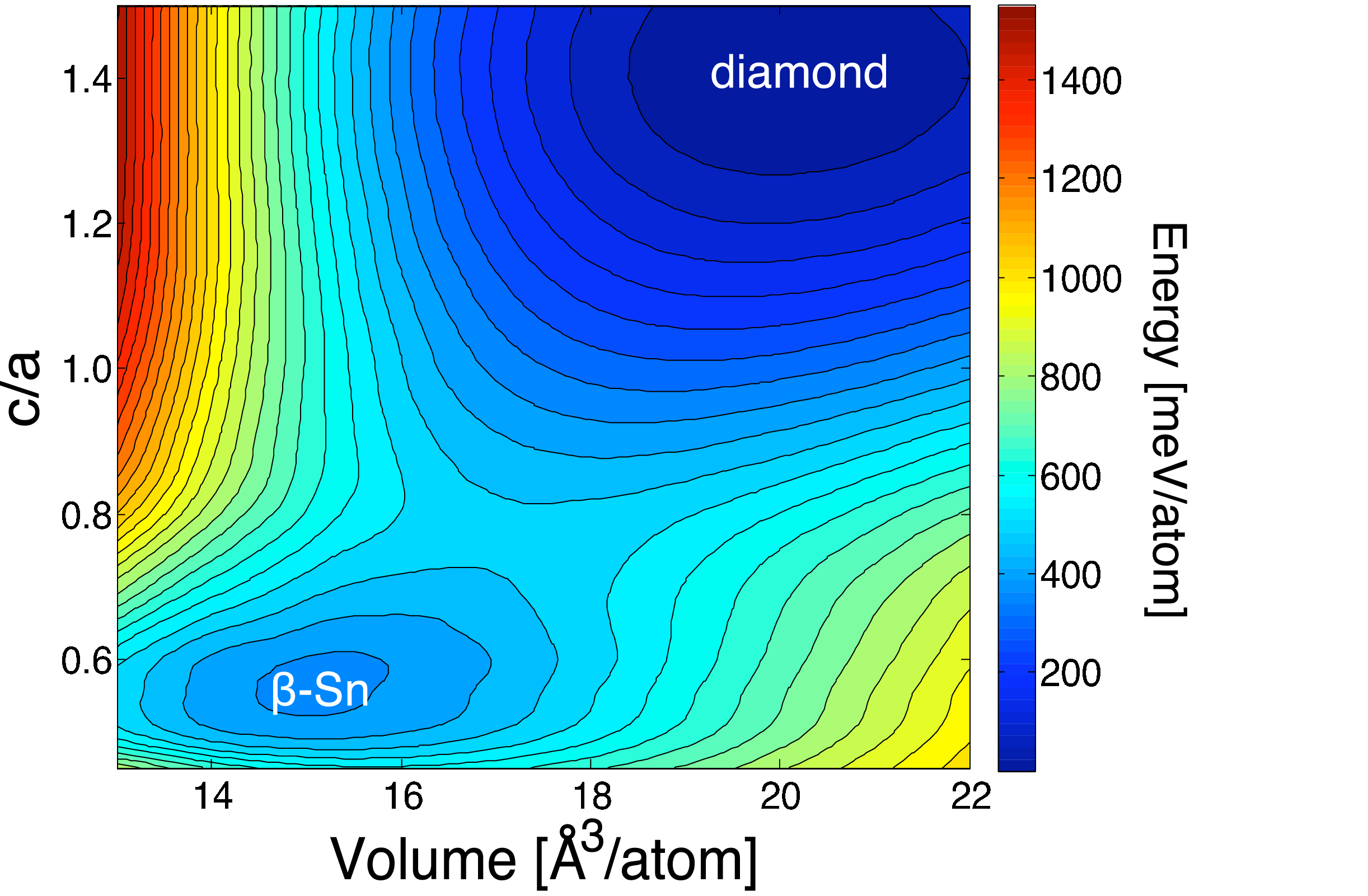}
  \caption{(color online) Energy landscape of Si as a function of volume and $c/a$ ratio for the HSE06 functional.  The two minima correspond to the diamond and $\beta$-Sn phases.  For the chosen unit cell, the $c/a$ ratio of the diamond phase is given by $\sqrt{2}$ instead of 1.  The contour lines are shown for every 50~meV/atom.}
  \label{fig:energylandscape}
\end{figure} 

To understand the trend of the various semi-local exchange-correlation functionals we note that a similar energy ordering for the various functionals and an agreement between the DMC and HSE06 results also occurs for Si single interstitial defects.~\cite{Batista06} The interstitial atom and its neighboring atoms have an increased coordination number of five or six.  For both the Si interstitial structures and the $\beta$-Sn phase, the increased coordination results in a more homogeneous charge density compared to the diamond phase.  This increased homogeneity explains why the LDA functional underestimates the energy difference between the two Si phases and the Si interstitial formation energies, respectively.  The lack of any gradient terms in LDA energetically favors more homogeneous charge density distributions.~\cite{Svendsen96} GGA functionals aim to correct this shortcoming of LDA.  However, PBE and PW91 violate the gradient expansion of Svendsen and von Barth~\cite{Svendsen96} for slowly varying density systems. Both functionals have second-order expansion coefficients that are too large.  One might therefore expect PBE to overestimate the effect of inhomogeneity. However, this analysis neglects the effect of higher order contributions.  We observe that the GGA functionals PBE and PW91 indeed improve the agreement with DMC, HSE06 and experiments but the resulting energy difference between the phases is still too small.   The more recent GGA functionals, WC, AM05 and PBEsol, which are designed to improve the description of solids, result in energy differences between the phases at or even below the LDA value and do not work for the transition pressure.  The gradient corrections alone appear insufficient.

\begin{figure}[tbh]
  \centering
  \includegraphics[width=7cm, angle=270]{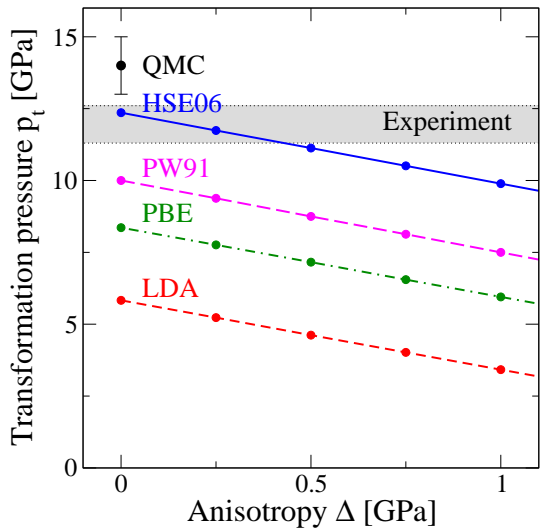}
  \caption{(color online) Effect of stress anisotropy on phase transition pressure.  The phase transformation from the diamond to the $\beta$-Sn phase of Si is reduced by anisotropic stress along the $[001]$ direction. The transition pressure reduction is similar for the various exchange-correlation functionals. For a stress anisotropy $\Delta$ along the $[001]$ direction compared to the other two cubic directions, the transition pressure is reduced by $2.4\,\Delta$.}
  \label{fig:anisotropy}
\end{figure}

Recent GW calculations by Rinke {\it et al.}~\cite{Rinke09} showed that the failure of LDA and GGA functionals for the interstitial formation energies in Si is due to underestimation of the vertical electron affinities of the interstitial defect configurations and related to the band-gap problem.  One might expect that this underestimation of the band gap in LDA and GGA also affects the accuracy of LDA and GGA functionals for the semiconductor to metal transition in Si.  Along a path in configuration space going from the diamond to the $\beta$-Sn phase, the band gap closes and the density of states at the Fermi level increases.  The bandgap closure occurs too early for the LDA and GGA functionals compared to HSE06 and the density of states at the Fermi level is higher for the LDA and GGA functionals as well.  This difference in groundstate properties predicted by the LDA and GGA functionals correlate with an energy difference along the path that is too low for the LDA and GGA functionals.   The inclusion of short-ranged Hartree-Fock exchange in the hybrid HSE06 functional is optimized to describe covalent bonds and improves the band gap in semiconductors.~\cite{Heyd05}  This may be the reason for the good agreement of the HSE06 results for the transition pressures in Si with experiments and for the interstitial formation energies with GW calculations~\cite{Rinke09} and DMC.~\cite{Batista06}

\subsection{Effect of stress anisotropy on the transition pressure}

The predicted transition pressure in DMC is at the upper end of the experimental values. There are various possibilities for why the DMC transition pressure prediction is somewhat high.  DMC might overestimate the energy difference between the metallic $\beta$-Sn phase and the semiconducting diamond structure due to the fixed-node error of the trial wave function.  In order to obtain the transition pressure accurate to within 1~GPa, one has to determine the energy difference between the two phases to within 20~meV/atom.  Because of the large difference in the electronic structure of the two phases, metallic versus semiconducting, it is possible that a sufficiently accurate cancellation of fixed-node error between the two phases does not occur.

Another possibility is that the measured transition pressure could be affected by kinetic effects and stress anisotropies in the diamond anvil cell experiments.~\cite{Hu86, Gupta80} Kinetic or hysteresis effects typically lead to an increase in transition pressure over the thermodynamic equilibrium transition pressure.  Stress anisotropy in the sample can lower the transition pressure by stabilizing the $\beta$-Sn phase over the diamond phase.~\cite{Cheng01, Gaal-Nagy06}

We determine the effect of non-hydrostatic stresses on the transition pressure following the approach by Cheng {\it et al.}~\cite{Cheng01}  The phase transformation from the diamond to the $\beta$-Sn phase occurs when
\begin{equation}
  \label{eq:transition}
  F_2 - F_1 + W \le 0,
\end{equation}
where $W$ is the work done by the system during the phase transformation and $F_1$ and ${F_2}$ are the energies of the diamond and the $\beta$-Sn phase, respectively.  In our analysis we neglect the effect of stress anisotropy on the zero-point energy and finite temperature phonon entropy contributions to the free energy and simply include those corrections as determined by Ga\'al-Nagy {\it et al.}~\cite{Gaal-Nagy99}  We assume a uniform non-hydrostatic compression along the cubic and tetragonal axes of the two crystal structures with a stress tensor, $\sigma$, of the form
\begin{equation}
  \sigma = 
  \left (
    \begin{array}[c]{c c c}
      \scriptstyle p-\Delta/3 & \scriptstyle 0 & \scriptstyle 0 \\
      \scriptstyle 0 & \scriptstyle p-\Delta/3 & \scriptstyle 0 \\
      \scriptstyle 0 & \scriptstyle 0 & \scriptstyle p+2\Delta/3 \\
    \end{array}
  \right ),
\end{equation}
where $p$ is the average applied stress and $\Delta$ is a measure of the stress anisotropy between the x or y axes and the z-axis.  For this loading condition, the work $W$ is
\begin{equation}
  \label{eq:work}
  W = p_x \int_{(1)}^{(2)} l_y\, l_z \, dl_x + p_y \int_{(1)}^{(2)} l_z\, l_x \, dl_y + p_z \int_{(1)}^{(2)} l_x\, l_y \, dl_z,
\end{equation}
where $p_x$, $p_y$ and $p_z$ are the three diagonal entries of the stress tensor $\sigma$, and $l_x$, $l_y$ and $l_z$ are the lattice constants of the crystal structures.  The virtual work under non-hydrostatic loading is path dependent~\cite{Wang95} so we calculate the work along the shortest path, following Ref.~\onlinecite{Cheng01}.

To determine the initial and final point of the path integral for the virtual work, we calculate the energy $E(V, c/a)$ of the Si diamond and $\beta$-Sn phases as a function of volume and $c/a$ ratio using the LDA, PBE, PW91 and HSE06 functionals.  We leave out the WC, AM05 and PBEsol functionals since they significantly underestimate the transition pressure.  We use a body-centered tetragonal unit cell for the simulations with the three lattice vectors $(\sqrt{2}\,a_0, \sqrt{2}\,a_0,0)$, $(\sqrt{2}\,a_0, -\sqrt{2}\,a_0,0)$ and $(0,0,c_0)$.  In these lattice vectors, the $c/a$ ratio of the diamond phase is $\sqrt{2}$ instead of one.  Figure~\ref{fig:energylandscape} illustrates the resulting energy landscape for the HSE06 functional.  The two minima correspond to the diamond and $\beta$-Sn phases of Si.  From the energy landscape $E(V, c/a)$, we determine the lattice parameters of the two phases as a function of applied stress tensor $\sigma$ using the relations: \begin{eqnarray}
  \label{eq:pressures}
  p_x &=& - \frac{\partial E}{\partial V} + \frac{c/a}{V} \frac{\partial E}{\partial (c/a)} \\
  p_z &=& - \frac{\partial E}{\partial V} - 2 \cdot \frac{c/a}{V} \frac{\partial E}{\partial (c/a)}.
\end{eqnarray}

Equations~\ref{eq:transition} and ~\ref{eq:work} determine the equilibrium transition pressures as a function of stress anisotropy $\Delta$.  Figure~\ref{fig:anisotropy} compares the resulting transition pressures for the LDA, PBE and HSE06 functionals with the range of experimental values as a function of the applied stress anisotropy $\Delta$.  All three functionals predict a similar behavior.  The anisotropy in the loading condition $\Delta$ linearly reduces the transition pressure by $2.4\, \Delta$.  The linear coefficient is similar for all three functionals and close to the values determined previously for the LDA and PW91 functional.~\cite{Cheng01, Gaal-Nagy06} Considering that the mechanical strength of silicon under uniaxial compression at ambient conditions is about 7~GPa,~\cite{Peterson82} it may be reasonable to assume that deviatory stresses of the order of 0.5~GPa could be present in the diamond-anvil cell experiments with a pressure medium and even larger deviatory stresses without a pressure medium.  This would lower the observed transition pressure by 1.2~GPa compared to perfect hydrostatic compression.  This strong influence of the stress anisotropy may explain the range of experimentally observed transition pressures, particularly the differences in the diamond anvil cell experiments with and without a pressure medium.  The effect of stress anisotropy may also explain the difference between the experimental values and the DMC prediction.

\section{Conclusion}
\label{sec:conclusion}

We performed calculations of the transition pressure for the high-pressure phase transformation in Si from the semiconducting diamond to the metallic $\beta$-Sn phase.  Comparisons with experimental values benchmark the accuracy of DMC methods and various exchange-correlation functionals in DFT.  The hybrid functional HSE06 and the DMC method predict similar transition pressures with values of 12.4 and $14.0\pm 1.0$~GPa, respectively, while semilocal LDA and GGA functionals predict lower transition pressures ranging from 3.5 to 10.0~GPa.

The DMC transition pressure is slightly above the experimental range of values of 11.3-12.6~GPa while the HSE06 functional agrees with the experiments.  The DMC energies could be affected by fixed-node error which could be reduced using the backflow transformation.~\cite{Rios06} The LDA prediction of 5.8~GPa is too low.  The GGA functionals PBE and PW91 improve the prediction but still underestimate the pressure.  The more recently developed GGA functionals, WC, AM05 and PBEsol, perform worse than the PBE and PW91 functionals and predict transition pressures more similar to the LDA.  Comparison with DMC and GW calculations for point defects in Si indicate that the underestimation of the transition pressure may be related to the underestimation of the bandgap in Si.

Calculations for anisotropic loading conditions show that the experimental transition pressure could be affected by stress anisotropy.  Stress anisotropy can dramatically reduce the transition pressure for the diamond to $\beta$-Sn transformation.  An anisotropy of only 0.5~GPa along any of the cubic axes reduces the transition pressure by 1.2~GPa, explaining the range of transition pressures observed in the diamond anvil cell experiments with and without a pressure medium and possibly also the difference between the experiments and the DMC.

\section{Acknowledgements}

The work was supported by the U.S. Department of Energy under Contract Nos.\ DE-FG02-99ER45795 and DE-FG05-08OR23339 and the National Science Foundation under Contract Nos.\ EAR-0703226, EAR-0530813 and DMR-0908653.  This research used computational resources of the National Energy Research Scientific Computing Center, which is supported by the Office of Science of the U.S. Department of Energy under Contract No.\ DE-AC02-05CH11231, the National Center for Computational Sciences at Oak Ridge National Laboratory, which is supported by the Office of Science of the U.S. Department of Energy under Contract No.\ DE-AC05-00OR22725, the National Center for Supercomputing Applications under grant DMR050036, the Ohio Supercomputing Center and the Computation Center for Nanotechnology Innovation at Rensselaer Polytechnic Institute.  We thank Derek Warner, John Perdew, Dario Alf{\`e}, Ann Mattsson and Art Ruoff for helpful discussions.

%

\end{document}